\documentclass[letterpaper,12pt]{article}
\usepackage{graphicx}
\usepackage{color,amssymb,enumerate,float,amsmath}
\usepackage{tikz}
\usetikzlibrary{matrix}
\usepackage{caption}
\usepackage{subcaption}

\usepackage{ifpdf}
\ifpdf
	\usepackage[pdftex,unicode,implicit]{hyperref}

	\hypersetup{
  	pdftitle     = {}, 
  	pdfkeywords  = {},
  	pdfauthor    = {},
  	pdfcreator   = {pdf\LaTeXe\ with package \flqq hyperref\frqq},
  	pdfproducer  = {pdf\LaTeXe\ with package \flqq hyperref\frqq},
  	pdfpagemode  = UseNone,  
  	pdffitwindow = true,  
  	unicode      = true,
  	plainpages   = true,
  	colorlinks   = true,  
  	citecolor    = black,  
  	urlcolor     = blue, 
  	linkcolor    = black
	}

\else

  \usepackage[dvips]{graphicx}
  \usepackage[unicode,implicit]{hyperref}

\fi 

\makeatletter
\@addtoreset{equation}{section}
\makeatother

\begin{document}
	
\thispagestyle{empty}

\begin{center}
{\bf \LARGE Exact   analytic solutions  in $2+1$ Ho\v rava gravity with cosmological constant }
\vspace*{15mm}

{\large Jorge Bellorin} $^{1,a}$
{\large  Claudio B\'orquez}$^{2,b}$
{\large and Byron Droguett}$^{1,c}$
\vspace{3ex}

$^1${\it Department of Physics, Universidad de Antofagasta, 1240000 Antofagasta, Chile.}

$^2${\it Facultad de Ingenier\'ia, Universidad San Sebasti\'an, Lago Panguipulli 1390, Puerto Montt, Chile.
}

\vspace{3ex}
$^a${jorge.bellorin@uantof.cl}, \quad
$^b${\tt claudio.borquez@uss.cl},\quad
$^c${\tt 
 byron.droguett@uantof.cl}\hspace{.5em}

\vspace*{15mm}
{\bf Abstract}
\begin{quotation}{\small\noindent
}
We investigate the static solutions with rotational symmetry in the nonprojectable Ho\v rava theory in \(2+1\) dimensions. We consider all  inequivalent terms of the effective theory, including the cosmological constant. We find two distinct types of solutions: the first one corresponds to a Lifshitz solution, while the second one is obtained through a coordinate transformation of the equations of motion, and exhibits Lifshitz scaling only asymptotically.

\end{quotation}
\end{center}

\thispagestyle{empty}

\newpage

\section{Introduction}
General relativity on a $(2+1)$-dimensional spacetime has no dynamical degrees of freedom, unlike the ($3+1$)-dimensional case, where gravitational interactions are mediated by the two traceless tensor modes. In $(2+1)$ dimensions, the curvature of spacetime is trivial or constant if the presence of the cosmological constant is considered, making it a topological theory. Despite this, nontrivial effects of global origin arise, such as the Bañados-Teitelboim-Zanelli (BTZ) black hole with a negative cosmological constant \cite{Banados:1992wn}. 

General relativity is a nonrenormalizable theory in the perturbative framework beyond one loop \cite{tHooft:1974toh, Goroff:1985th}. Furthermore, issues such as dark energy and dark matter motivate the exploration of alternative theories of gravity. In this context, Ho\v rava gravity emerges as a candidate to complete the high-energy regime of general relativity \cite{Horava:2009uw, Horava:2008ih}. The theory introduces an anisotropy between space and time, which reduces the full diffeomorphism gauge symmetry and explicitly breaks the Lorentz symmetry. This symmetry reduction generically gives rise to an additional local scalar mode, which persists even in $(2+1)$ dimensions and plays a central role in mediating local gravitational interactions. The theory is formulated on a foliation of spatial hypersurfaces along a preferred time direction, which must be preserved by the reduced diffeomorphism group. A suitable set of field variables to describe this preferred foliation is given by the Arnowitt-Deser-Misner (ADM) variables \cite{Arnowitt:1962hi}. The reduced symmetry permits the inclusion of terms of high-order in spatial derivatives, while keeping the time derivatives fixed at second order. Moreover, this symmetry enables the formulation of two versions of the Ho\v rava theory: the projectable and nonprojectable versions \cite{Horava:2009uw}. In particular, the nonprojectable case is a theory with second-class constraints. For comprehensive reviews about Ho\v rava gravity, see \cite{Barvinsky:2023mrv, Wang:2017brl}.

In this work, we study static and rotationally symmetric solutions of the $(2+1)$-dimensional nonprojectable Ho\v{r}ava gravity with cosmological constant. We work within the low-energy regime. The potential includes an additional term compared to the ADM formulation of general relativity (and coupling constants may differ from their values in general relativity). The additional term depends on spatial derivatives of the lapse function, and it plays a crucial role in ensuring the consistency of the theory \cite{Blas:2009qj}.

Exact analytical solutions of Ho\v{r}ava gravity are in general difficult to find. In the large-distance limit, the additional term of the Lagrangian leads to additional terms in the field equations. Indeed, in $(3+1)$ dimensions with a cosmological constant, the static and spherically symmetric solutions of the nonprojectable theory are known only numerically \cite{Bellorin:2015oja}. Our aim in this work is to get a better comprehension of this class of solutions in Ho\v rava gravity. Our strategy to obtain the exact analytical solutions is to lower the dimensionality to $(2+1)$ dimensions. We obtain simplifications on the field equations, as we show in the next section, which eventually allows us to obtain the analytic solutions. Therefore, the $(2+1)$ model may give important insights into the structure of this kind of solutions.

There are several examples where the $(2+1)$ field equations have helped to understand the physical features of the Ho\v{r}ava theory. The well-known case of the rest-particle solution of $(2+1)$ general relativity  \cite{Deser:1983tn} determines a flat geometry with a deficit angle. This solution is used as reference for defining the asymptotically flat condition \cite{Ashtekar:1993ds}. A different solution with cosmological constant was found in \cite{Deser:1983nh}. In nonprojectable Ho\v{r}ava theory (without cosmological constant), despite the fact that the field equations are different, the same flat conical solution is obtained \cite{Bellorin:2019zho}. Several kinds of exact and analytic solutions of the $(2+1)$-dimensional theory were studied in Ref.~\cite{Shu:2014eza}. Among them, a Lifshitz space was found. We find a similar behavior in our study, as we comment below. Another case is the critical Ho\v{r}ava theory, defined by the condition of the coupling constant of the kinetic term in the Lagrangian takes the value $\lambda =1/d$, where $d$ is the spatial dimension. This case exhibits conformal symmetry in the kinetic term. The additional mode disappears, leaving the same number of degrees of freedom as in general relativity. Hence, in $(2+1)$ dimensions the critical Ho\v{r}ava theory has no local degrees of freedom. Nevertheless, in a previous study on the $(2+1)$ critical theory without cosmological constant \cite{Bellorin:2020ixa}, we found both flat and non-flat solutions. This is in contrast to $(2+1)$ general relativity, where all solutions are necessarily flat.

On quantum grounds, the study of the $(2+1)$ dimensional case provides useful insights for understanding the quantization and renormalization. In the nonprojectable version of Ho\v rava gravity, a detailed analysis of the cancellation of dangerous divergences was carried out in \cite{Bellorin:2022qeu}, with special usage of the $(2+1)$ theory.

In $(3+1)$-dimensional nonprojectable Ho\v{r}ava theory, one of the static and spherically symmetric solutions that naturally arises is a wormhole \cite{Kiritsis:2009vz, Bellorin:2014qca} (more precisely, a solution with a throat). This solution was previously found \cite{Eling:2006df} in the Einstein-Aether theory \cite{Jacobson:2000xp}, which is dynamically equivalent to the large-distance limit of nonprojectable Ho\v{r}ava gravity under the condition of hypersurface orthogonality \cite{Jacobson:2010mx,Jacobson:2013xta}. In the approach used in Ref. \cite{ Bellorin:2014qca} to obtain this solution explicitly, a key step is to perform a change of coordinate motivated by a quadratic structure present in the field equations. The new coordinate provides a clear identification of the critical point where the throat is located. The same kind of configuration was studied in Ho\v rava theory with cosmological constant \cite{Bellorin:2015oja}. In this case, the solution with the throat was found numerically. The other solution is a monotonically growing naked singularity, without throat or horizon. Additionally, black holes and solutions with universal horizons have been studied \cite{Barausse:2011pu,Blas:2011ni,Sotiriou:2014gna,Basu:2016vyz, Rubio:2023eva}. In addition, an extra $U(1)$ symmetry has been included \cite{Horava:2010zj,Zhu:2011xe}. Similar solutions have been found in this model \cite{Greenwald:2010fp, Greenwald:2011ca,Lin:2016myf}.

As we have commented, the $(2+1)$-dimensional case leads to a tractable system of field equations under the conditions of staticity, rotational symmetry (and vanishing of the shift vector, a condition that we also use). The first solution we find arises just by direct substitution of a power-law ansatz in the field equations. We give in advance that this solution represents a kind of Lifshitz geometry, in the sense that there is an anisotropic scaling between the radial component of the spatial metric and the lapse function, which can be interpreted as the time component of an underlying spacetime metric. Similar solutions with Lifshitz anisotropy have arisen in Ho\v{r}ava gravity with cosmological constant in the context of holography, see for example \cite{Griffin:2011xs,Griffin:2012qx,Shu:2014eza}. Moreover, they were used for anisotropic holography previously to the formulation of the Ho\v{r}ava theory \cite{Kachru:2008yh}. They have been interpreted as holographic duals of anisotropic Lifshitz models of matter fields. The component of the metric along the extra dimension of the bulk exhibits an anisotropic scaling with respect to the time component. In this sense, they are geometrically similar to the solution we find here. We also remark that another interesting feature of our solution is that it exhibits the Lifshitz scaling exactly point to point, whereas the analogous $(3+1)$-dimensional solution has Lifshitz scaling only asymptotically \cite{Bellorin:2015oja}.

We find a second exact solution under the same geometric assumptions by performing a change of the radial coordinate similar to the procedure done in $(3+1)$ dimensions that we have commented. We find that there is an analogous quadratic structure formed by the field equations, and in this case we can manage to obtain the solution in terms of the new coordinate explicitly. Hence, this solution is also an analytic solution. We remark that the procedure is rather nontrivial, in the sense that the original system of field equations is very involved. Besides the previous solution that can be find by using an ansatz, other solutions are difficult to find analytically. Given the relationship between the old and the new radial coordinate, we may explore whether there is critical point that can be interpreted as the throat of a wormhole.

\section{Solution of the field equations}
\subsection{The field equations}
Ho\v rava theory is based on a foliation of spacetime with an absolute physical character. This foliation consists of space-like hypersurfaces with a privileged time direction. A suitable set of  variables to describe the foliation are the ADM variables
\begin{equation}	
g_{\mu\nu} =
\left( {\begin{array}{cc}
	-N^{2}+N_{k}N^{k} & N_j\\
	N_i & g_{ij} \\
	\end{array} } \right) \,,
 \label{metricg}
\end{equation}
where $N$ is the lapse function, $N_{k}$ is the shift function and $g_{ij}$ is the spatial metric (spatial indices are raised and lowered with $g_{ij}$).
The diffeomorphisms that preserve the foliation form a subset of the general diffeomorphism group of general relativity. This leads to spatial and temporal coordinates having different scaling \([t]=-z\) , \([x]=-1\), where $z$ represents the anisotropic scaling factor. The transformations of coordinates and the ADM fields are given by:
\begin{equation}
\begin{array}{l}
\delta t = f(t) \,,
\hspace{2em}
\delta x^i = \zeta^i(t,\vec{x}) \,,
\\[1ex]
\delta N = \zeta^k \partial_k N + f \dot{N} + \dot{f} N \,,
\\[1ex]
\delta N_i = \zeta^k \partial_k N_i + N_k \partial_i \zeta^k 
            + \dot{\zeta}^j g_{ij} + f \dot{N}_i + \dot{f} N_i \,,
\\[1ex]
\delta g_{ij} = \zeta^k \partial_k g_{ij} + 2 g_{k(i} \partial_{j)} \zeta^k 
                + f \dot{g}_{ij}  \,.
\end{array}                
\label{fdiff}
\end{equation}
The  most general potential  that contribute to large distances in $2+1$ dimensions is 
\begin{equation}
 \mathcal{V} = - \beta R - \alpha a_k a^k +\Lambda \,,
 \label{potential}
\end{equation}
where $\beta$ and $\alpha$ are coupling constants, $\Lambda$ is a cosmological constant and the covariant vector is defined by
\begin{equation}
 a_i = \frac{ \partial_i N }{ N }\,.
\end{equation}
The effective action for large distance of  Ho\v{r}ava gravity has the form
\begin{equation}
	S = 
	\frac{1}{2\kappa} \int dt d^2x 
	\sqrt{g} N \left( K_{ij}K^{ij} - \lambda K^2 
	+ \beta R + \alpha a_k a^k-\Lambda \right) 
 \,,
	\label{Eaction}
\end{equation}
where  the extrinsic curvature is
\begin{equation}
 K_{ij} = 
 \frac{1}{2N} \left( \dot{g}_{ij} - 2 \nabla_{(i} N_{j)} \right) \,,
 \qquad 
 \label{extrinsiccurvature}
\end{equation} 
its trace $K \equiv g^{ij} K_{ij}$, $\kappa$ and $\lambda$ are coupling constants and the dot stands for derivative with respect to $t$, $\dot{g}_{ij} \equiv \partial g_{ij} / \partial t$. 

The field equations of the ADM variables are obtained by varying the action (\ref{Eaction}) with respect to the fields. This yields
\begin{eqnarray}
	&& 
	 K^{ij}K_{ij} - \lambda K^{2} + \beta R + \alpha a_i a^i -\Lambda 
     - 2 \alpha \frac{\nabla^{2} N}{N} 
	 =
0
\,,
	\label{deltaN}
	\\ && 
	G^{ijkl} \nabla_{j} K_{kl}  
	=
	0
 \,,
	\label{deltaNi}
	\\ 
	&& 
	\frac{1}{\sqrt{g}} \frac{\partial}{\partial t} 
	  \left( \sqrt{g} G^{ijkl} K_{kl} \right) 
	+ 2 G^{mikl} \nabla_m ( N^j K_{kl} ) 
	- G^{ijkl} \nabla_{n}( N^{n} K_{kl} )  \nonumber 
	\\ &&
	+ 2 N (K^i{}_k K^{jk} - 2 \lambda KK^{ij}) 
	- \frac{1}{2} N g^{ij} G^{klmn} K_{kl} K_{mn} 
	- \beta \left( \nabla^{ij} N - g^{ij} \nabla^{2}N \right)
	\nonumber 
	\\ &&
	+ \frac{\alpha}{N} \left(\nabla^i N \nabla^j N 
	     - \frac{1}{2} g^{ij} \nabla_{k} N \nabla^{k} N \right)
	-\beta N \left( R^{ij} - \frac{1}{2} g^{ij} R \right) -\frac{1}{2}\Lambda N g^{ij}
	\nonumber 
	\\ &&
	=
	0
 \,,
	\label{deltag}
\end{eqnarray}
where $G^{jikl} \equiv \frac{1}{2} \left( g^{ik} g^{jl} + g^{il} g^{jk} \right) - \lambda g^{ij} g^{kl}$. 

We consider the gauge $N_i = 0$ for simplicity and the static case. With these conditions, the extrinsic curvature is zero and therefore there is no contribution of the parameter $\lambda$ to the kinetic term. Since we have two spatial dimensions, the identity $R_{ij} - \frac{1}{2} g_{ij} R = 0$ holds. Hence, the Eq. (\ref{deltag}) takes the form
\begin{equation}
   	\beta 
    \left( 
    \nabla_{ij} N - g_{ij} \nabla^2 N 
    \right)
  	- \frac{\alpha}{N} 
    \left(
    \nabla_i N \nabla_j N 
 	- \frac{1}{2} g_{ij} \nabla_{k} N \nabla^{k} N 
    \right)+\frac{1}{2}\Lambda N g_{ij}
  	= 0 \,.
   \label{eq1metric}
\end{equation}
The trace of the Eq. (\ref{eq1metric}), assuming $\beta \neq 0$, yields 
\begin{equation}
	\nabla^2 N=\gamma N \,,
	\label{nablaN}
\end{equation}
where\footnote{In the conventions we are using, the cosmological constant is $-\gamma = -\Lambda/\beta$.} $\gamma=\Lambda/\beta$. The Eq. (\ref{deltaN}) together with Eq. (\ref{nablaN}) is reduced to 
\begin{eqnarray}
	 \beta R + \alpha a_i a^i -\gamma\left(
     \beta 
     + 2 \alpha 
     \right)
     &=&
	0 \,.
	\label{deltaNreduced}
	\end{eqnarray}
We replace the Eqs. (\ref{nablaN}) and (\ref{deltaNreduced}) in Eq. (\ref{eq1metric}), obtaining
 
	\begin{equation}
	    \beta\nabla_{ij}N-\alpha N^{-1}\nabla_iN\nabla_jN+g_{ij}N\left(
        \alpha\gamma-\frac{1}{2}\beta R
        \right)=0 
        \,.
    \label{eq1metricmodf}
	\end{equation}
    
In this investigation we introduce the ansatz of a static spatial metric and lapse function $N$ with rotational symmetry, which in polar coordinates is 
\begin{eqnarray}
    \,dS^2&=&f^{-1}(r)\,dr^2+r^2\,d\theta^2\,,
\end{eqnarray}
hence under this ansatz the Eqs.(\ref{nablaN}) and (\ref{deltaNreduced}) are, respectively,
 \begin{eqnarray}
N^{''}+\Big(\frac{1}{2}\frac{f^{'}}{f}+\frac{1}{r}\Big)N^{'}-\gamma \frac{N}{f}
\label{Eq1}
&=&
0\,,
\\
\alpha f \left(
N^{'}
\right)^2+\beta \frac{f^{'}}{r}N^2-\left(
\beta+2\alpha
\right)\gamma N^2
&=&
0\,,
\label{Eq2}
\end{eqnarray}
where the prime indicates derivative with respect to $r$.
The Eq. (\ref{eq1metricmodf}) generates two  equations, the $rr$ and $\theta\theta$ component, the other equations are identically zero,
\begin{eqnarray}
    \beta N^{''} +\frac{\beta f^{'}}{2rf}
    \left(
    -N+rN^{'}
    \right)+\alpha\gamma\frac{N}{f}-\alpha\frac{
    \left(
    N^{'}
    \right)^2}{N}
    &=&0\,,
    \label{Eq3}
    \\
   - \frac{\beta}{2}\frac{f^{'}}{f}+\beta\frac{N^{'}}{N}+\alpha\gamma \frac{r}{f}
    &=&
    0\,.
    \label{Eq4}
\end{eqnarray}
We have a system of equations (\ref{Eq1}-\ref{Eq4}) which is redundant. The equations of first order in derivatives  (\ref{Eq2}, \ref{Eq4}) imply the equations of second order in derivatives (\ref{Eq1}, \ref{Eq3}). Therefore, the system (\ref{Eq2}, \ref{Eq4}) is consistent for the unknown functions $N(r)$  and $f(r)$. When \(\alpha = 0\) and \(\Lambda \neq 0\), we recover the system of field equations corresponding to \(2+1\) general relativity under the assumptions of staticity, rotational symmetry and a vanishing shift vector.

In the most general case ($\alpha , \Lambda \neq 0$), it is convenient to combine the Eqs. (\ref{Eq2}) and (\ref{Eq4}), resulting in the following quadratic form
\begin{equation}
    \left(
    r\frac{N^{'}}{N}
    \right)^2+2\frac{\beta}{\alpha}
    \left(
    r\frac{N^{'}}{N}
    \right)
    =
    \frac{\Lambda}{\alpha}\frac{r^2}{f}\,,\qquad \alpha\neq 0\,.
    \label{quadraticform}
\end{equation}
In the particular case where $\Lambda=0$, the Eq. (\ref{quadraticform}) can be written as
\begin{equation}
    \left(
    r\frac{N^{'}}{N}
    \right)\left(
    r\frac{N^{'}}{N}+2\frac{\beta}{\alpha}
    \right)=0\,.
\end{equation}
This equation has two different solutions: the first one occurs when \( r\frac{N^{'}}{N} = 0 \), which implies that the lapse function and \( f \) are constants. Consequently, the solution corresponds to the flat case. The second solution is determined by
\begin{equation}
    r\frac{N^{'}}{N}+2\frac{\beta}{\alpha}=0\,,
\end{equation}
which can be solved by integration. The constant of integration $r_{0}$ in the $f$ function can not be absorbed. Therefore the exact solution is given by
\begin{eqnarray}
    \,ds^2&=&
    -r^{-4\beta/\alpha}\,dt^2+
    \left(
    \frac{r}{r_0}
    \right)^{-4\beta/\alpha}\,dr^2+r^2\,d\theta^2\,.
\end{eqnarray}
The functions $N$ and $f$ are singular at infinity or zero, depending on the sign of the factor $\beta/\alpha$. This solution coincides with the one obtained in the critical case \cite{Bellorin:2020ixa}.

\subsection{First exact Lifshitz  solution}
\begin{figure}[H]
    \begin{minipage}[b]{0.8\linewidth}
    \centering
\includegraphics[width=.8\linewidth]{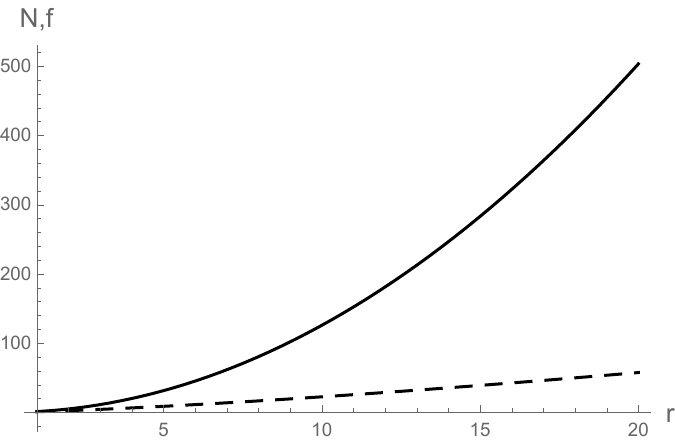} 
    \vspace{4ex}
  \end{minipage}
  \caption{The solid and dashed lines represent the Lifshitz-type functions \( f \) and \( N \), respectively, as given in Eq. (\ref{Lif-Sol}). The parameter values \( \alpha = -4 \), \( \beta = 3 \), and \( \Lambda = 1 \) have been used for this representation (the signs of $\beta$ and $\Lambda$ correspond to a negative cosmological constant). }
  \label{Fig1}
\end{figure}

The first solution we find arises under a power-law ansatz for $N$ and $f$. These fields take the form
\begin{equation}
    N=r^a\,,\qquad f=cr^b, \qquad c>0\,.
    \label{Lif-Sol}
\end{equation}
When we evaluate this anzat in the field equations (\ref{Eq2}, \ref{Eq4}), we get
\begin{eqnarray}
&&r^{2 a} \left(c r^{b-2} \left(\alpha  a^2+b \beta \right)-\gamma  (2 \alpha +\beta )\right)=0\,,
\label{eq2asym}
\\
&&
\frac{\beta  (2 a-b)}{2 r}+\frac{\alpha  \gamma  r^{1-b}}{c}=0\,.
\label{eq4asym}
\end{eqnarray}
All equations imply that \( b = 2 \) for all values of \( a \) and \( c \). Moreover, by considering Eqs. (\ref{eq2asym}) and (\ref{eq4asym}), we obtain the exact solutions \( (a_+, c_-) \) and \( (a_-, c_+) \), where
\begin{eqnarray}
a_\pm&=&
-\frac{\beta}{2\alpha^2} 
\left(
\pm\sqrt{8 \alpha ^2+4 \alpha  \beta +\beta ^2}+2 \alpha +\beta \right)\,, 
\\
c_\mp&=&
\frac{\gamma}{2 \beta (\alpha +2 \beta )}
\left[
\beta 
\left(
\beta \mp\sqrt{8 \alpha ^2+4 \alpha  \beta +\beta ^2}
\right)+2 \alpha ^2+2 \alpha  \beta
\right]\,.
\end{eqnarray}
The condition \( c > 0 \) imposes constraints on the parameters of the theory. The parameter \( a_\pm\) can take either positive or negative values.  
The obtained solution corresponds point-to point to an exact Lifshitz solution. This Lifshitz solution in $(2+1)$ dimensions differs from the solution obtained in the $(3+1)$ dimensional nonprojectable Ho\v rava gravity in the infrared regime, where the solution approaches the Lifshitz geometry only asymptotically \cite{Bellorin:2015oja}. In this case, the solution corresponds to a Lifshitz geometry in the whole space and time. As shown in Figure~\ref{Fig1}, the functions $N$ and $f$ exhibit monotonic behavior. The coupling constants were selected to guarantee that the parameter \(c\) remains positive. A positive value of \(a\) was chosen to ensure that the solutions exhibit a monotonically increasing behavior. If the power $a_{\pm}$ coincides with the anisotropic scaling factor \(z\), then we recover vacuum Lifshitz-type solutions with the same anisotropic scaling of the solutions found in Refs. \cite{Griffin:2011xs, Griffin:2012qx}. The spatial Ricci scalar of this solution is equal to
\begin{equation}
 R = 2c_{\mp} \,.
\end{equation}
Hence, over a spatial surface at a given time, this is a solution of constant curvature. Moreover, $c_{\mp} > 0$, which implies that the spatial curvature is positive. As usual, one may build a $(2+1)$ spacetime by incorporating the lapse function as the timelike component of the metric. In this case, the $(2+1)$ Ricci scalar also takes a constant value, $^{(3)} R = 2 c_{\mp} ( 1 + a_{\pm} + a_{\pm}^2 )$.

Substituting \(\alpha = 0\) into Eqs.~(\ref{eq2asym}) and (\ref{eq4asym}), the parameters take the values \(a = 1\), \(b = 2\), and \(c = \gamma/2\). Note that this requires \(\gamma > 0\). The corresponding metric then takes the form:
\begin{equation}
    ds^2 = -r^2\,dt^2 + \frac{2}{\gamma r^2}\,dr^2 + r^2\,d\theta^2\,,
\end{equation}
which corresponds to the limit of zero angular momentum and zero mass of the BTZ solution \cite{Banados:1992wn}.

 \subsection{Second solution}
\subsubsection{The coordinate transformation}
The structure of the field equations suggests a transformation in the radial coordinate. We first analyze a general transformation on the field equations (\ref{Eq1}--\ref{Eq4}). Therefore, we introduce a new radial coordinate \(\xi\) that helps to understand the structure of the field equations. In terms of this new radial coordinate, the spacetime metric can be expressed as
\begin{eqnarray}
    \,ds^2&=&-N^2(\xi)\,dt^2+h(\xi)\,d\xi^2+r^2(\xi)\,d\theta^2
    \,.
    \label{newmetric}
\end{eqnarray}
To find the factors we require, we utilize the chain rule
  \begin{eqnarray}
      \frac{1}{r}\left(r\frac{N^{'}}{N}\right)\frac{dr}{d\xi}&=&\frac{1}{N}\frac{dN}{d\xi}\,,
      \label{change1}
      \\
      \frac{f^{'}}{f}
      &=&
     \left( \frac{\alpha}{\Lambda}\frac{f}{ r^2}\right)^{-1}\frac{\,d}{\,d\xi}\left(\frac{\alpha}{\Lambda}\frac{f}{ r^2}\right)\frac{\,d\xi}{\,dr}+\frac{2}{r}\,,
    \label{change2}
  \end{eqnarray}
  we recall that the prime indicates derivate with respect to $r$.
The Eq. (\ref{Eq1}) can be explicitly written in terms of the new variable. For this, we perform the following manipulations
  \begin{eqnarray}
N^{''}+\Big(
\frac{1}{2}\frac{f^{'}}{f}+\frac{1}{r}
\Big)N^{'}-\gamma \frac{N}{f}
&=&
\frac{1}{fr^2}\left(r^2fN^{'}\right)^{'}-\Big(
\frac{1}{2}\frac{f^{'}}{f}+\frac{1}{r}
\Big)N^{'}-\gamma \frac{N}{f}\,,
\label{nablagammaconvform}
\end{eqnarray}
then, by replacing the Eqs. (\ref{change1}) and (\ref{change2}) into the Eq. (\ref{nablagammaconvform}), we have 
\begin{eqnarray}
  && 
    \frac{\,d}{\,d\xi}
    \left[
    \left(
    \frac{\alpha}{\Lambda}\frac{f}{ r^2}
      \right)
    \left(
    r\frac{N^{'}}{N}
    \right)
    \right]
    -\frac{1}{2}
    \left(
    r\frac{N^{'}}{N}
    \right)\frac{\,d}{\,d\xi}
    \left(
    \frac{\alpha}{\Lambda}\frac{f}{ r^2}
      \right)
      \nonumber
      \\
      &&
    + \frac{1}{r}\left(
    \frac{\alpha}{\Lambda}\frac{f}{ r^2}
    \right)
    \left[
    \left(
    r\frac{N^{'}}{N}
    \right)^2
    +\left(
    r\frac{N^{'}}{N}
    \right)
    - \frac{\alpha}{\beta}\left(
    \frac{\alpha}{\Lambda}\frac{f}{ r^2}
    \right)^{-1}
    \right] \frac{\,dr}{\,d\xi}
    =0\,,
\end{eqnarray}
thus, we obtain a general formula for a radial coordinate transformation
\begin{eqnarray}
    \frac{\,dr}{\,d\xi}&=& -\frac{ \frac{\,d}{\,d\xi}
    \left(
    \left(
    \frac{\alpha}{\Lambda}\frac{f}{ r^2}
      \right)
    \left(
    r\frac{N^{'}}{N}
    \right)
    \right)
    -\frac{1}{2}
    \left(
    r\frac{N^{'}}{N}
    \right)\frac{\,d}{\,d\xi}
    \left(
    \frac{\alpha}{\Lambda}\frac{f}{ r^2}
      \right)}{\left(
    \frac{\alpha}{\Lambda}\frac{f}{ r^2}
    \right)
    \left(
    \left(
    r\frac{N^{'}}{N}
    \right)^2
    +\left(
    r\frac{N^{'}}{N}
    \right)
    - \frac{\alpha}{\beta}\left(
    \frac{\alpha}{\Lambda}\frac{f}{ r^2}
    \right)^{-1}
    \right) }r\,.
    \label{drdxi}
\end{eqnarray}
The lapse function in the new coordinate is determined by the Eq. (\ref{change1})

\begin{eqnarray}
    \frac{1}{N}\frac{\,d N}{\,d\xi}&=&
    -\frac{ \frac{\,d}{\,d\xi}
    \left(
    \left(
    \frac{\alpha}{\Lambda}\frac{f}{ r^2}
      \right)
    \left(
    r\frac{N^{'}}{N}
    \right)
    \right)
    -\frac{1}{2}
    \left(
    r\frac{N^{'}}{N}
    \right)\frac{\,d}{\,d\xi}
    \left(
    \frac{\alpha}{\Lambda}\frac{f}{ r^2}
      \right)}{\left(
    \frac{\alpha}{\Lambda}\frac{f}{ r^2}
    \right)
    \left(
    \left(
    r\frac{N^{'}}{N}
    \right)^2
    +\left(
    r\frac{N^{'}}{N}
    \right)
    - \frac{\alpha}{\beta}\left(
    \frac{\alpha}{\Lambda}\frac{f}{ r^2}
    \right)^{-1}
    \right) }
    \left(r\frac{N^{'}}{N}\right)
    \,,
    \label{Nsol}
\end{eqnarray}
and the new radial component of the spatial metric is given by

\begin{eqnarray}
    h(\xi)&=&\frac{1}{f}\left(\frac{\,dr}{\,d\xi}\right)^2\,.
    \label{Rsol}
\end{eqnarray}

We now perform the specific coordinate transformation.
The Eq. (\ref{quadraticform}) has a quadratic form, which we will analyze for the general case  $\alpha, \Lambda,\beta-1 \neq 0$. For this purpose, we complete the square and obtain the following expression:
 \begin{eqnarray}
     \left[\left(
     1+\frac{\beta}{\alpha}r\frac{N^{'}}{N}
     \right)^2
     +\chi
     \left(
     \frac{\beta}{\alpha}r\frac{N^{'}}{N}
     \right)^2\right]
     &=&
    \rho
     \,,
     \label{analiticform}
 \end{eqnarray}
 where
 \begin{eqnarray}
     \chi&=&
     \left(\frac{\alpha}{\beta}
     \right)^2-1
     \,,
     \\
     \rho&=& 
      1+\frac{\Lambda}{\alpha}\frac{r^2}{f}
      \,.
      \label{parameters}
 \end{eqnarray}
The Eq. (\ref{analiticform}) can be parameterized using trigonometric or hyperbolic functions, depending on the sign of \(\chi\) and \(\rho\). First, we focus on the case \(\chi > 0\) and \(\rho > 0\), which leads to the following trigonometric change of variable:
 \begin{equation}
     \left(1+\frac{\beta}{\alpha}r\frac{N^{'}}{N}\right)
     =
    \sqrt{\rho} \sin{\xi}
      \,,
     \qquad
     \sqrt{\chi}
     \left(
     \frac{\beta}{\alpha}r\frac{N^{'}}{N}
     \right)
     =
    \epsilon \sqrt{\rho}\cos{\xi}
    \,,
    \label{sincos}
 \end{equation}
 where $\xi\in (-\pi/2,\pi/2)$ and $\epsilon=\pm 1$. The following factors can be written explicitly in terms of $\xi$:
 \begin{eqnarray}
     r\frac{N^{'}}{N}&=&
     \frac{\alpha}{\beta}\left(
     \epsilon\sqrt{\chi}\tan\xi-1
     \right)^{-1}
     \label{Indpef1}
     \,,
     \\
     \frac{\alpha}{\Lambda}\frac{f}{ r^2}
     &=&
     \frac{\left(\sqrt{\chi } \tan (\xi )-1\right)^2}{2 \sqrt{\chi } \tan (\xi )+\chi -1}\,.
     \label{Indpef2}
 \end{eqnarray}
As a consequence, Eqs. (\ref{drdxi}--\ref{Rsol}) can be solved by integration, allowing us to obtain exact analytic solutions. The homogeneous first-order spatial partial differential equation (\ref{drdxi}) has the explicit form\footnote{In this work we consider the case $\epsilon=+1$.}
 \begin{eqnarray}
\frac{1}{r}\frac{\,dr}{\,d\xi}
&=&
\frac{\beta\sqrt{\chi}\sec ^2\xi  
\left( 
1-\sqrt{\chi}  \tan \xi
\right)}{
\left(
2 \sqrt{\chi } \tan\xi+\chi 
-1\right)
\left(
-\alpha +\beta  \sqrt{\chi } \tan\xi +\beta 
\chi 
\right)}\,.
\label{difequation}
\end{eqnarray}
This expression, according to Eq.~(\ref{Indpef1}), is not defined at 
\( \xi = \arctan\left( \frac{1}{\sqrt{\chi}} \right) \). As a result, the differential equation (\ref{difequation}) does not admit any critical points. In fact, the derivative vanishes only at \( \xi = \arctan\left( \frac{1}{\sqrt{\chi}} \right) \), which coincides with a singularity and therefore is outside the domain of the solution. Additionally, the points  \(\xi = \arctan\left( \frac{1 - \chi}{2\sqrt{\chi}} \right)\) and
\(\xi = \arctan\left( \frac{\alpha - \beta\chi}{\beta\sqrt{\chi}} \right)
\) are outside the solution domain. 

By integrating Eq. (\ref{difequation}), we obtain the exact solution for the radial function
 \begin{eqnarray}
    r(\xi )&=& 
    c_1\left|
    2 \sqrt{\chi } \tan \xi +\chi -1
    \right|^{\frac{\beta  (\chi +1)}{2 (-2 \alpha +\beta( \chi +1) )}}
    \left|
     \beta  \chi -\alpha +\beta  \sqrt{\chi } \tan \xi 
    \right|^{- \left(
    \frac{-\alpha +\beta(1+\chi)}{-2 \alpha +\beta( \chi +1) }
    \right)}\,,
    \nonumber
    \\
    &&
    \label{analyticsolr}
 \end{eqnarray}
the absolute value present in the radial function guarantees that the function is well-defined. This solution, along with Eq. (\ref{Indpef1}), provides the exact solution for the lapse function
 \begin{eqnarray}
     N(\xi)&=&
     c_2\left|
     \frac{-1+\chi+2\sqrt{\chi}\tan\xi}{\alpha-\beta\chi-\beta\sqrt{\chi}\tan\xi}
     \right|^{\frac{\alpha}{2\alpha-\beta(1+\chi)}}\,,
 \end{eqnarray}
where \(c_1\) and \(c_2\) are constants of integration. The function \(f\) can be determined using Eqs. (\ref{Indpef2}) and (\ref{analyticsolr}),
\begin{eqnarray}
    f&=&
    \frac{\Lambda c_1^2}{\alpha}
    \frac{\left(
     \sin\xi- \frac{1}{\sqrt{\chi}}\cos\xi
     \right)^2}{1-\left(
     \sin\xi- \frac{1}{\sqrt{\chi}}\cos\xi
     \right)^2}
    \frac{\left|-1+\chi+2 \sqrt{\chi } \tan \xi
    \right|^{\frac{\beta  (\chi +1)}{ (-2 \alpha +\beta( \chi +1) )}}}{
    \left| \beta  \chi -\alpha +\beta  \sqrt{\chi } \tan \xi
    \right|^{2 
    \left(
    \frac{-\alpha +\beta(1+\chi)}{-2 \alpha +\beta( \chi +1) }
    \right)}}\,,
\end{eqnarray}
and the  new radial component (\ref{Rsol}) has the form 
\begin{eqnarray}
h(\xi)&=&
\frac{\alpha  \beta ^2 \chi  \sec ^4\xi }{\Lambda \left(
2 \sqrt{\chi } \tan \xi +\chi -1
\right) 
\left(
-\alpha +\beta  \chi+\beta  \sqrt{\chi } \tan \xi 
\right)^2}\,.
\end{eqnarray}

To analyze the invariants of the theory, we use the structure of the Riemann tensor, which in 2 spatial dimensions takes the following form:
\begin{equation}
R_{ijkl} = \frac{R}{2} \left( g_{ik} g_{jl} - g_{il} g_{jk} \right)\,.
\end{equation}
Therefore, it is sufficient to study the scalar curvature. It results 
    \begin{eqnarray}
R &=&
\frac{1}{r(\xi)} \frac{\,d\xi}{\,dr} \frac{\,df(\xi)}{\,d\xi}
\nonumber \\
  &=& 
  \frac{2 \Lambda \left[ \alpha \chi - \beta \chi^2 + \beta + \sqrt{\chi} (\alpha - 2 \beta (\chi + 1))\tan \xi \right]}{\alpha \beta \left( 2 \sqrt{\chi} \tan \xi + \chi - 1 \right)}
  \,.
\end{eqnarray}
We observe that the curvature scalar has a singularity when the new coordinate takes the value
\begin{equation}
    \xi = \arctan\left( \frac{1 - \chi}{2 \sqrt{\chi}} \right)\,,
\end{equation}
This point corresponds to the origin $r=0$. Thus, we find that the scalar curvature diverges at this point, indicating the presence of a physical singularity at the origin. 

\begin{figure}[H]
    \begin{minipage}{0.7\linewidth}
    \centering
\includegraphics[width=.8\linewidth]{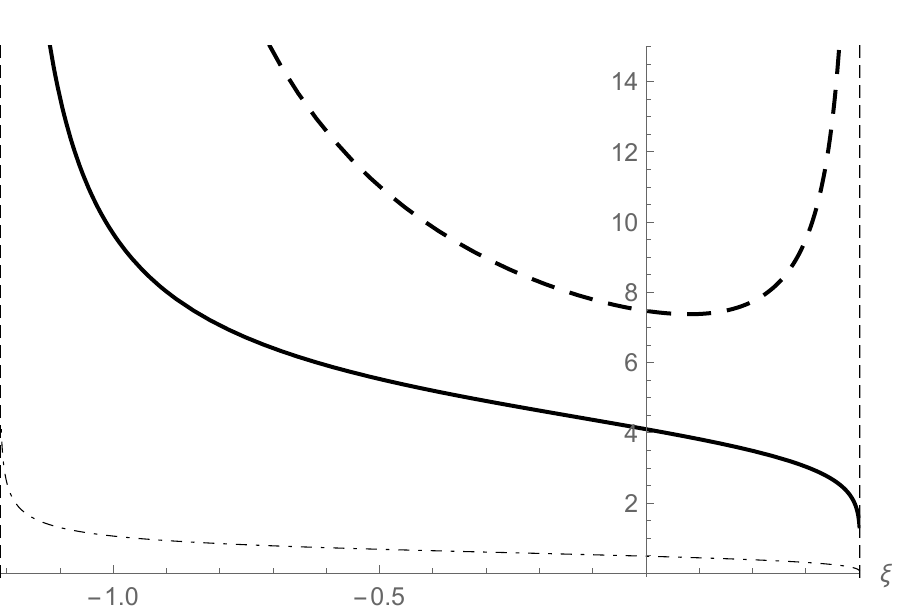} 
    \vspace{4ex}
  \end{minipage}
  \caption{The solid, dashed, and dot-dashed lines represent the functions \( r \), \( f \), and \( N \), respectively. The left asymptote is located at \(\arctan\left(\frac{\alpha - \beta \chi}{\beta \sqrt{\chi}}\right)\), while the right asymptote is at \(\arctan\left(\frac{1 - \chi}{2\sqrt{\chi}}\right)\).  
For this representation, the parameter values \( \alpha = -4 \), \( \beta = 3 \), \( \Lambda = 1 \), \( c_1 = 14 \), and \( c_2 = 0.05 \) have been used (the signs of $\beta$ and $\Lambda$ correspond to a negative cosmological constant).  }
  \label{Fig2}
\end{figure}

In Fig. \ref{Fig2}, we present the components of the static spatial metric with rotational symmetry, along with the lapse function. The solid, dashed, and dotted lines correspond to the functions \( r \), \( f \), and \( N \), respectively.  
We observe that the function \( r \) grows monotonically in the inverse direction of $\xi$, similar to the lapse function \( N \). In contrast, the function \( f \) decreases in this direction near \( r = 0 \), until it reaches a minimum (as a function of $\xi$). We observe that asymptotically (as $\xi$ goes to the left), the functions $N$ and $f$ grow monotonically. For $r>0$ the solution is well defined for all $r\in (0,\infty)$. The functions equivalent to the spacetime metric components in the coordinate $r$, $N^2(r)$ and $f^{-1}(r)$, are different form zero, continuous and positive. Hence, there are no critical points in the original coordinate $r$, which implies that there are no horizons (or physical singularities) in $r\in (0,\infty)$. Therefore, the singularity present at the origin is a naked singularity. 

\begin{figure}[H]
\begin{minipage}[b]{0.7\linewidth}
    \centering
    \includegraphics[width=0.9\linewidth]{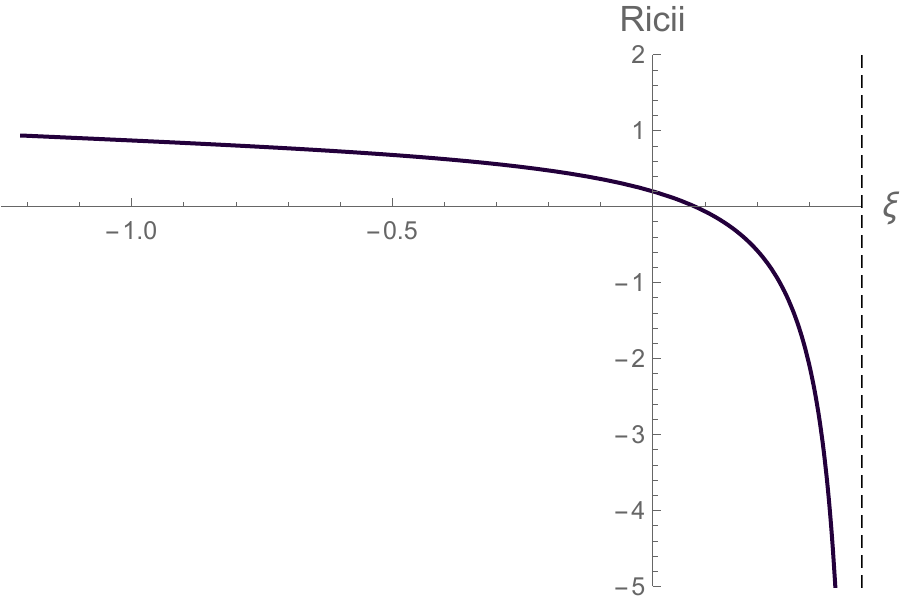} 
    \vspace{4ex}
\end{minipage}
\caption{The solid line represents the Ricci scalar as a function of \( \xi \), with the right asymptote located at \(\arctan\left(\frac{1 - \chi}{2\sqrt{\chi}}\right)\). This representation is based on the parameter values \( \alpha = -4 \), \( \beta = 3 \), and \( \Lambda = 1 \).
}
\label{Fig3}
\end{figure}

The plot in Fig. \ref{Fig2} shows that the dominant asymptotic modes for these functions are of different scale. Hence, this solutions acquires Lifshitz scaling asymptotically. In Fig. \ref{Fig3} we plot the scalar curvature, with a pole at the origin. The behavior of the functions agrees with the numerical integration of the Eqs. (\ref{Eq2}, \ref{Eq4}), where $r$ is the independent variable.

 Second, we focus on the case \(\chi < 0\) and \(\rho > 0\), which leads to the following hyperbolic change of variable
 \begin{equation}
     \left(1+\frac{\beta}{\alpha}r\frac{N^{'}}{N}\right)
     =
    \sqrt{\rho} \cosh{\xi}
      \,,
     \qquad
     \sqrt{-\chi}
     \left(
     \frac{\beta}{\alpha}r\frac{N^{'}}{N}
     \right)
     =
    \epsilon \sqrt{\rho}\sinh{\xi}
    \,.
    \label{sihncohs}
 \end{equation}
  Again, the following factors can be written explicitly in terms of $\xi$:
 \begin{eqnarray}
     r\frac{N^{'}}{N}&=&
     \frac{\alpha}{\beta}\left(
     \sqrt{-\chi}\coth\xi-1
     \right)^{-1}
     \label{Indpef1H}
     \,,
     \\
     \frac{\alpha}{\Lambda}\frac{f}{ r^2}
     &=&
     \frac{\left(\sqrt{-\chi } \coth (\xi )-1\right)^2}{2 \sqrt{-\chi } \coth(\xi )+\chi -1}\,.
     \label{Indpef2H}
 \end{eqnarray}
As a consequence, Eqs. (\ref{drdxi}--\ref{Rsol}) can be solved by integration, allowing us to obtain exact analytic solutions. The homogeneous first-order spatial partial differential equation (\ref{drdxi}) has the explicit form
 \begin{eqnarray}
\frac{1}{r}\frac{\,dr}{\,d\xi}
&=&
\frac{\beta  \text{csch}^2\xi  \left(
\chi  \coth \xi +\sqrt{-\chi }
\right)}{\left(
2 \sqrt{-\chi } \coth \xi +\chi -1
\right) \left(
\alpha -\beta  \sqrt{-\chi } \coth \xi -\beta  \chi \right)}\,.
\label{difequationhyperbolic}
 \end{eqnarray}
This expression, according to Eq.~(\ref{Indpef1H}), is not defined at \( \xi = \text{arccoth}\left( 1/\sqrt{-\chi} \right) \). As a result, the differential equation (\ref{difequationhyperbolic}) does not admit any critical points. In fact, the derivative vanishes only at \( \xi = \text{arccoth}\left( 1/\sqrt{-\chi} \right) \), which coincides with a singularity, and therefore is outside the domain of the solution. Additionally, the points $\xi = \text{arccoth}\left(\frac{1-\chi}{2\sqrt{-\chi}} \right)$ and $\xi = \text{arccoth}\left(\frac{\alpha - \beta\chi}{\beta\sqrt{-\chi}} \right)$ are outside the solution domain.

By integrating Eq. (\ref{difequation}), we obtain the exact solution for the radial function
\begin{eqnarray}
    r (\xi ) &=&
    c_{1}\sqrt {
    \left|\frac {\cosh \xi - 1}{\sinh  \xi  }
    \right|}
    \frac{
    \left|
    \sqrt {-\chi} \left(
    1 + \tanh^{2} 
    \frac{\xi}{2}
    \right)
    + \left(
    \chi - 1
    \right)
    \tanh  \frac{\xi}{2} 
    \right| ^{\frac {\beta\, \left( \chi + 1 \right) }{2\beta\,(\chi + 1) -4\,\alpha}}} 
    {\left| 
    \beta\sqrt {-\chi}
    \left(
    1 + \tanh^{2} \frac{\xi}{2}
    \right) 
    + 2\left(
    \beta\,\chi - \alpha
    \right)\tanh  \frac{\xi}{2} 
    \right| ^{\frac{\beta\,(\chi + 1) -\alpha}{\beta\,(\chi + 1) - 2\,\alpha}}}
    \,,
    \nonumber
    \\
\end{eqnarray}
and the lapse and $f$ functions are given by

\begin{eqnarray}
    N (\xi ) &=&
    c_{2}\, 
    \left|
    \frac{ \beta\sqrt {-\chi} 
    \left( 1 + \tanh^{2}  \frac{\xi}{2} 
    \right)
    + 2\left(
    \beta\,\chi - \alpha
    \right)\tanh \frac{\xi}{2} }{ \sqrt {-\chi} 
    \left( 
    1 + \tanh^{2}  \frac{\xi}{2}   \right) 
    + \left(
    \chi - 1
    \right) \tanh \frac{\xi}{2}  }
    \right|^{\frac {\alpha}{\beta\,(\chi + 1) - 2\,\alpha}}
    \,,
\end{eqnarray}

\begin{eqnarray}
    f(\xi) &=&
    c_{1}^{2}\frac{\Lambda}{\alpha}
    \frac{\left(\sqrt{-\chi } \coth (\xi )-1\right)^2}{2 \sqrt{-\chi } \coth(\xi )+\chi -1}
    r^2(\xi)
    \,.
\end{eqnarray}
We have introduced two coordinate transformations depending on the sign of $\chi$, which allows us to cover all possible ranges of the constants $\alpha$ and $\beta$.

\section{Conclusions}

In this work, we have analyzed static solutions with rotational symmetry and vanishing shift function of the nonprojectable Ho\v{r}ava theory in $(2+1)$ dimensions with cosmological constant. We have based the study on the effective action for large-distances, considering all the inequivalent terms in the Lagrangian.

We have found two new exact solutions. The first one represents an exact Lifshitz configuration for all points of space. It has positive constant curvature, in the sense of the two-dimensional spatial surfaces. We catalog this as a Lifshitz geometry due to the anisotropic degree of growing between the lapse function and the (inverse of) the radial component of the spatial metric. We believe that it is an interesting feature that solutions with Lifshitz scaling arise naturally as vacuum solutions of the Ho\v{r}ava theory when a cosmological constant is present. As we have commented previously, in $(3+1)$ dimensions a similar (numerical) solution is known with asymptotic Lifshitz scaling, and, in holographic contexts, solutions with anisotropic scalings on the extra dimension of the bulk has been used as holographic duals of Lifshitz matter fields on the boundary. These spaces are vacuum solutions of the Ho\v{r}ava theory. Here we have found this solution with a similar behavior, in the context of static rotationally symmetric configurations.

The second solution is obtained after performing a coordinate transformation on the radial direction. A combination of the field equations acquires a quadratic structure. The coordinate transformation is done on the basis of this quadratic equation, using trigonometric or hyperbolic functions to solve it. Actually, these are two different solutions, but we refer them as one, understanding that one is the hyperbolic counterpart of the trigonometric one. Several functional combinations that arise in the field equations, and the original radial coordinate itself, can be solved explicitly in terms of the new coordinate. This allows us to determine all the metric components by integration, obtaining an exact and analytical solution. The differential equation yielding the coordinate transformation indicates that there are no critical points, consequently, no wormhole is formed. Qualitatively, we observe that this solution acquires Lifshitz scaling asymptotically. As $r\rightarrow\infty$, the dominant modes of $N^2$ and $f^{-1}$ are of different scale. The solution exhibits a naked singularity. Indeed, the analogous solution in $(3+1)$ dimensions (with or without a cosmological constant) is also a naked singularity, since one of the asymptotic halves of the wormhole exhibits a naked singularity at the boundary. The $(2+1)$ solution we have found repeats this  behavior, in this case without a throat.

We have verified that both solutions coincide with the direct numerical integration of the original field equations (\ref{Eq2}) and (\ref{Eq4}),  where the original radial coordinate is taken as the independent variable. We believe that having succeeded in obtaining the exact analytic solutions of the Ho\v{r}ava gravity, considering the complexity of the field equations, is an important step. We hope that these solutions can be used to further understanding the dynamics of this theory, and also can be extended by relaxing some of the assumptions we have used in this study.

\end{document}